\documentclass[a4paper,12pt,preprint]{elsarticle}

\usepackage{lineno,hyperref}
\modulolinenumbers[5]

\journal{SoftwareX}

\usepackage{hyperref}
\usepackage{graphicx,subcaption}
\usepackage{placeins}
\usepackage{dcolumn}
\usepackage{bm}
\usepackage{mathtools} 

\usepackage[utf8]{inputenc}
\usepackage[T1]{fontenc}
\usepackage{float}

\usepackage{color}

\definecolor{LinkColor}{RGB}{48, 67, 163}

\usepackage{listings}
\newcommand{\code}[1]{\texttt{\lstinline{#1}}} 

\newcommand{\MJOLNIR}{\texttt{MJOLNIR}}
\newcommand{\MJOLNIRGui}{\texttt{MJOLNIRGui}}
\newcommand{\QQQ}{\mathbf{Q}}
\newcommand{\kkk}{\mathbf{k}}
\newcommand{\PPP}{\mathbf{P}}
\usepackage{standalone}
\usepackage{tikz}
\usetikzlibrary{shapes.callouts,calc,arrows,shapes,positioning,shapes,arrows,fadings,decorations.pathreplacing,shadows}
\usetikzlibrary{fadings,through,angles,quotes,decorations.pathmorphing,patterns}
\tikzset{snake it/.style={decorate, decoration=snake}}
\usepackage{ifthen}
\newcommand{\CC}{C\nolinebreak\hspace{-.05em}\raisebox{.4ex}{\tiny\bf +}\nolinebreak\hspace{-.10em}\raisebox{.4ex}{\tiny\bf +}}
\def\CC{{C\nolinebreak[4]\hspace{-.05em}\raisebox{.4ex}{\tiny\bf ++}}}

\begin{document}

\begin{frontmatter}

\title{MJOLNIR: A Software Package for Multiplexing Neutron Spectrometers}


\author[psi,nbi,coor]{Jakob Lass}
\author[psi]{Henrik Jacobsen}
\author[psi]{Daniel G. Mazzone}
\author[nbi]{Kim Lefmann}

\address[psi]{Laboratory for Neutron Scattering and Imaging, Paul Scherrer Institut, CH-5232 Villigen, Switzerland}
\address[nbi]{Nanoscience Center, Niels Bohr Institute, University of Copenhagen, DK-2100 Copenhagen \O , Denmark}
\cortext[coor]{\textit{Correspondence to}: 5232 Villigen PSI, Switzerland. \\ \textit{E-mail address}: jakob.lass@nbi.ku.dk}

\begin{abstract}
Novel multiplexing triple-axis neutron scattering spectrometers yield significant improvements of the common triple-axis instruments. While the planar scattering geometry keeps ensuring compatibility with complex sample environments, a simultaneous detection of scattered neutrons at various angles and energies leads to tremendous improvements in the data acquisition rate. Here we report on the software package \MJOLNIR~that we have developed to handle the resulting enhancement in data complexity. Using data from the new CAMEA spectrometer of the Swiss Spallation Neutron Source at the Paul Scherrer Institut, we show how the software reduces, visualises and treats observables measured on multiplexing spectrometers. The software package has been generalised to a uniformed framework, allowing for collaborations across multiplexing instruments at different facilities, further facilitating new developments in data treatment, such as fitting routines and modelling of multi-dimensional data.
\end{abstract}


\end{frontmatter}

\section{Introduction}
Inelastic neutron scattering instruments allow detailed studies of the dynamical structure factor, $S(\QQQ,\omega)$, where  $\QQQ$ is a scattering vector in reciprocal space and $\hbar\omega$ = $\Delta E$ an energy transfer. One of the work horses of modern neutron scattering are triple-axis instruments (see Fig.\  \ref{fig:TripleAxisOverview}), because they typically have a high neutron flux and good energy resolution \cite{TripleAxis}.

Unfortunately, standard triple-axis instruments cover only a single ($\QQQ$, $\hbar\omega$)-position per acquisition time, which leads to isolated cuts along lines of $\omega$ or $\QQQ$ during experiments.
Multiplexing triple-axis instruments extend the triple-axis concept by employing multiple analysers and detectors. This allows for simultaneous measurement of large $S(\QQQ,\omega)$ areas, while preserving the high flux. This is in contrast to direct geometry time-of-flight mapping instruments, which have an even larger coverage of reciprocal space, but which pay the price in form of a significant reduction in incoming flux. 

Examples of multiplexing triple-axis instruments are  RITA-II \cite{Rita2}, FlatCONE \cite{FlatCone}, UFO \cite{UFO}, MultiFLEXX \cite{Groitl2016}, PUMA \cite{PUMA}, Bambus \cite{Lim2014}, SPINS, MACS \cite{MACS}, and CAMEA \cite{Lass2020CAMEA}. 
The point clouds measured on these spectrometers have to be treated in a fundamentally new fashion, therefore pushing the data reduction onto a new complexity level. In some sense the usage of these instruments are simplified through the reduction of movable parts and parameters while the post processing of data increases significantly in complexity, very similar to the case of direct-geometry time-of-flight spectrometers. 



Here, we present the software package \MJOLNIR{}, which has been primarily developed for the new secondary neutron spectrometer CAMEA at the Paul Scherrer Institut (PSI), Switzerland. However, the software can handle data from every multiplexing triple-axis neutron spectrometer, enabling utility across facilities. \MJOLNIR{} offers tools to quickly convert data from detector counts to reciprocal space, and to visualise the data in 1, 2 and 3 dimensions. The software has been written with focus on being user-friendly, offering a scripting, a command line, and a graphical interface. The software was successfully used during the commissioning phase of the CAMEA spectrometer\cite{Lass2020CAMEA} and subsequently for the data treatment of acquired data\cite{Janas2020,Allenspach2020}.

\MJOLNIR{} has been coded in the open source language Python and can be downloaded at PSI website\cite{PSICAMEADataTreatment}. \MJOLNIR{} makes use of renowned Python packages, such as matplotlib\cite{Matplotlib}, scipy\cite{Scipy}, numpy, and Pandas\cite{PandasSoftware}, and as such is radially distributable and compatible across platforms through Python's PyPi interface. We mention that a trend for academic software towards Python is currently recognisable. In fact, well-established Matlab\cite{MATLAB} programs such as SpinW\cite{Toth2015} are being rewritten to allow for Python bindings. New large scale facilities including the European Spallation Source \cite{Andersen2020} plan to run with Python-based systems\cite{Mantid14}. 

In this article we present the main features of \MJOLNIR{}. The full documentation including extensive tutorials is available at the PSI website\cite{PSICAMEADataTreatment}. Before presenting \MJOLNIR{} we first explain, standard triple axis spectrometers and their differences to multiplexing instruments. We then describe the different coordinate systems used for in the data treatment. Subsequently, we describe the structure of \MJOLNIR{}, followed by examples of some of its features.

\subsection{Triple axis spectrometers}
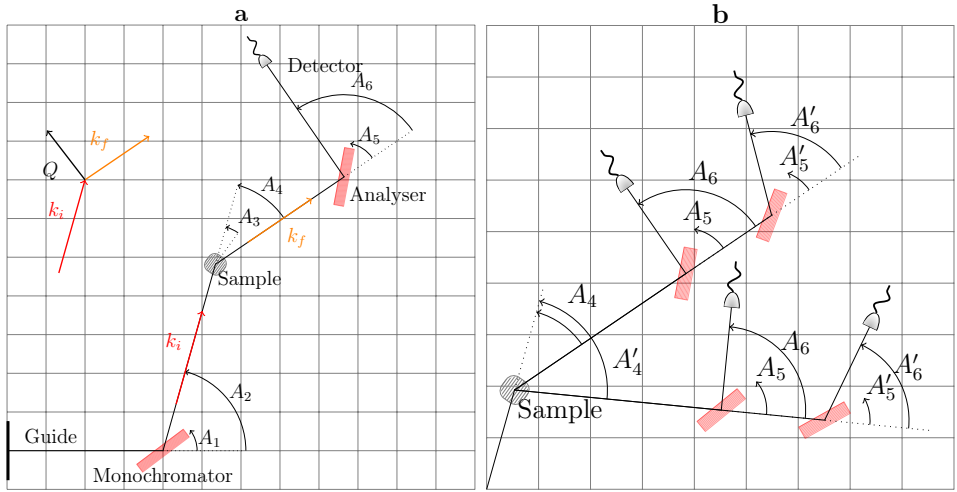
\begin{figure}[ht!]
    \centering
    \includestandalone[width=0.45\linewidth]{Figures/TripleAxisInstrument}
    \pgfmathsetmacro\Xa{0}%
\pgfmathsetmacro\Ya{0}%
\pgfmathsetmacro\Za{0}%

\pgfmathsetmacro\thetaM{37.2}%
\pgfmathsetmacro\thetaA{45}%
\pgfmathsetmacro\thetaATwo{60}%
\pgfmathsetmacro\thetaS{-20}%

\pgfmathsetmacro\localThetaM{\thetaM}%
\pgfmathsetmacro\localThetaS{\thetaS+2*\thetaM}%

\pgfmathsetmacro\localLength{0.2}%

\pgfmathsetmacro\La{0.8}%
\pgfmathsetmacro\Lb{0.4}%

\pgfmathsetmacro\Xb{\Xa+\La}%
\pgfmathsetmacro\Yb{\Ya}%
\pgfmathsetmacro\Zb{\Za}%
\pgfmathsetmacro\localBLengthx{\localLength*cos(\thetaM)}%
\pgfmathsetmacro\LocalBLengthy{\localLength*sin(\thetaM)}%
\pgfmathsetmacro\localBLengthX{2*\localLength*cos(\localThetaM-\thetaM)}%
\pgfmathsetmacro\LocalBLengthY{2*\localLength*sin(\localThetaM-\thetaM)}%


\pgfmathsetmacro\Xc{\Xb+cos(2*\thetaM)*\Lb}%
\pgfmathsetmacro\Yc{\Yb+sin(2*\thetaM)*\Lb}%
\pgfmathsetmacro\Zc{\Zb}%
\pgfmathsetmacro\localCLengthx{\localLength*cos(\localThetaS)}%
\pgfmathsetmacro\LocalCLengthy{\localLength*sin(\localThetaS)}%

\pgfmathsetmacro\localCLengthX{2*\localLength*cos(\localThetaS-\thetaS)}%
\pgfmathsetmacro\LocalCLengthY{2*\localLength*sin(\localThetaS-\thetaS)}%

\pgfmathsetmacro\Zd{\Zc}%

\pgfmathsetmacro\LMono{0.2}%
\pgfmathsetmacro\DMono{0.05}%

\pgfmathsetmacro\LSample{0.1}%
\pgfmathsetmacro\DSample{0.1}%

\pgfmathsetmacro\xa{0}%
\pgfmathsetmacro\ya{-0.01}%
\pgfmathsetmacro\za{0}%

\pgfmathsetmacro\xblocal{0}%
\pgfmathsetmacro\yblocal{0.03}%
\pgfmathsetmacro\zb{0}%
\pgfmathsetmacro\xb{sin(\localThetaM)*\xblocal + cos(\localThetaM)*\yblocal}%
\pgfmathsetmacro\yb{-cos(\localThetaM)*\xblocal + sin(\localThetaM)*\yblocal}%

\pgfmathsetmacro\xclocal{0}%
\pgfmathsetmacro\yclocal{-0.03}%
\pgfmathsetmacro\zc{0}%
\pgfmathsetmacro\xc{cos(\localThetaS)*\xclocal - sin(\localThetaS)*\yclocal}%
\pgfmathsetmacro\yc{sin(\localThetaS)*\xclocal + cos(\localThetaS)*\yclocal}%

\pgfmathsetmacro\xdlocal{0}%
\pgfmathsetmacro\ydlocal{-0.03}%
\pgfmathsetmacro\zd{0}%

\pgfmathsetmacro\xelocal{0}%
\pgfmathsetmacro\yelocal{0}%
\pgfmathsetmacro\ze{0}%

\pgfmathsetmacro\rad{0.005}%

\begin{tikzpicture}[
    scale=4,
    pile2/.style={->,thick, shorten <=0pt, shorten >=1.5pt},
    every node/.style={color=black},
    crystal/.style={red, opacity=0.4}, 
    sample/.style={rounded corners,gray}
    ]
\draw[step=0.2cm,gray, thin] (0.799,0) grid (2.601,1.8);    
\node at (1.7,1.845) {\textbf{b}};
    
\coordinate (a) at (\Xa,\Ya,\Za);    
\coordinate (b) at (\Xb,\Yb,\Zb);
\coordinate (bLy) at ($(b)+(\localBLengthx,\LocalBLengthy,0)$);
\coordinate (bLx) at ($(b)+(\LocalBLengthy,-\localBLengthx,0)$);
\coordinate (bB) at ($(b)+(2.1*\localLength,0,0)$);

\coordinate (c) at (\Xc,\Yc,\Zc);
\coordinate (cLy) at ($(c)+(\localCLengthx,\LocalCLengthy,0)$);
\coordinate (cLx) at ($(c)+(-\LocalCLengthy,\localCLengthx,0)$);
\coordinate (cB) at ($(c)+(1.05*\localCLengthX,1.05*\LocalCLengthY,0)$);

\draw[sample,shift={(c)},rotate={\localThetaS}] (-\LSample/2.0,\DSample/2.0) rectangle (\LSample/2.0,-\DSample/2.0);
\node[shift={($(c)+(7*\LSample,-3.3*\DSample)$)}] {Sample};
\begin{scope}[shift={(c)},rotate={\localThetaS}]
\clip[shift={(c)},rotate={\localThetaS},rounded corners]  (-\LSample/2.0,\DSample/2.0) rectangle (\LSample/2.0,-\DSample/2.0);
\foreach \y in {-5,...,5}{
    \draw[sample] (-\LSample/2.0,-\y*\DSample*0.1) -- (\LSample/2.0,-\y*\DSample*0.1);
}
\end{scope}

\draw[] (b) -- (c);

\draw[dotted,shift={(c)}] (0,0,0) -- (cB); 

\node at ($(c)+(2.25*\localCLengthx,2.25*\LocalCLengthy,0)$) {$A_4$};

\pgfmathsetmacro\localCLengthxM{\localLength*cos(\localThetaS-40)}%
\pgfmathsetmacro\LocalCLengthyM{\localLength*sin(\localThetaS-40)}%

\node at ($(c)+(2.25*\localCLengthxM,2.25*\LocalCLengthyM,0)$) {$A_4'$};

\pgfmathsetmacro\Ld{0.4}%

\foreach \i [evaluate=\i as \wedge using \i*0.3] in {1,2} {
	\pgfmathsetmacro\thetaS{-20*\i}
	
	\foreach \j [evaluate=\j as \analyzer using \j*10] in {1,2} {
		\pgfmathsetmacro\thetaA{55-\analyzer}
		\pgfmathsetmacro\localThetaA{\thetaA+2*\thetaS+2*\thetaM}%
	
		\pgfmathsetmacro\Lc{0.4+0.4*\j+0*\i}%
		\pgfmathsetmacro\Xd{\Xc+cos(2*\thetaS+2*\thetaM)*\Lc}%
		\pgfmathsetmacro\Yd{\Yc+sin(2*\thetaS+2*\thetaM)*\Lc}%
	
		\pgfmathsetmacro\LocalDLengthx{\localLength*cos(\localThetaA)}%
		\pgfmathsetmacro\LocalDLengthy{\localLength*sin(\localThetaA)}%

		\pgfmathsetmacro\LocalDLengthX{2*\localLength*cos(\localThetaA-\thetaA)}%
		\pgfmathsetmacro\LocalDLengthY{2*\localLength*sin(\localThetaA-\thetaA)}%

		\pgfmathsetmacro\Xe{\Xd+cos(2*\thetaA+2*\thetaS+2*\thetaM)*\Ld}%
		\pgfmathsetmacro\Ye{\Yd+sin(2*\thetaA+2*\thetaS+2*\thetaM)*\Ld}%
		\pgfmathsetmacro\Ze{\Zd}%

		\pgfmathsetmacro\LocalELengthx{\localLength*cos(\localThetaA+\thetaA)}%
		\pgfmathsetmacro\LocalELengthy{\localLength*sin(\localThetaA+\thetaA)}%

		\pgfmathsetmacro\xd{sin(\localThetaA)*\xdlocal + cos(\localThetaA)*\ydlocal}%
		\pgfmathsetmacro\yd{-cos(\localThetaA)*\xdlocal + sin(\localThetaA)*\ydlocal}%

		\pgfmathsetmacro\xe{-cos(\localThetaA-\thetaA)*\yelocal - sin(\localThetaA-\thetaA)*\xelocal}%
		\pgfmathsetmacro\ye{-sin(\localThetaA-\thetaA)*\yelocal + cos(\localThetaA-\thetaA)*\xelocal}%

		\coordinate (d) at (\Xd,\Yd,\Zd);
		\coordinate (dLy) at ($(d)+(\LocalDLengthx,\LocalDLengthy,0)$);
		\coordinate (dLx) at ($(d)+(\LocalDLengthy,-\LocalDLengthx,0)$);
		\coordinate (dB) at ($(d)+(1.05*\LocalDLengthX,1.05*\LocalDLengthY,0)$);

		\coordinate (e) at (\Xe,\Ye,\Ze);
		\coordinate (eLx) at ($(e)+(\LocalELengthx,\LocalELengthy,0)$);
		\coordinate (eLy) at ($(e)+(-\LocalELengthy,\LocalELengthx,0)$);

		\draw[crystal,shift={(d)},rotate={\localThetaA}] (-\LMono/2.0,\DMono/2.0) rectangle (\LMono/2.0,-\DMono/2.0);
		\begin{scope}[shift={(d)},rotate={\localThetaA}]
		\foreach \y in {-5,...,4}{
		    \draw[crystal] (-\LMono/2.0,-\y*\DMono*0.1) -- (\LMono/2.0,-\y*\DMono*0.1);
		}
		\end{scope}

		\begin{scope}
			\clip[shift={(e)},rotate={\localThetaA+\thetaA}]  (0,-0.061) rectangle (0.06,0.03);
			\draw[black,inner color=black!30!white,shift={(e)},rotate={\localThetaA+\thetaA}] (0,0) ellipse (0.06 and 0.03);
			\path[draw=black,shift={(e)},rotate={\localThetaA+\thetaA}] (0,-0.03)--(0,0.03);
		\end{scope}

		\path [draw=black,snake it,thick,shift={(e)},rotate={\localThetaA+\thetaA}] (0.06,0)--(0.18,0);

		\draw[] (c) -- (d);
		\draw[] (d) -- (e);
		\draw[dotted,shift={(d)}] (0,0,0) -- (dB); 

	\pgfmathparse{\j > 1 ? 1 : 0}
	\draw pic [draw,->,angle radius=0.7cm,] {angle = dB--d--dLy}; 

	\ifthenelse{\pgfmathresult=0}{\node[anchor=base] at ($(d)+(1.25*\LocalDLengthx,1.05*\LocalDLengthy,0)$) {$A_5$};\node[anchor=base] at ($(d)+(1.85*\LocalDLengthx,1.85*\LocalDLengthy,0)$) {$A_6$};} 
	{\node[anchor=base] at ($(d)+(1.25*\LocalDLengthx,1.05*\LocalDLengthy,0)$) {$A_5'$};\node[anchor=base] at ($(d)+(1.85*\LocalDLengthx,1.85*\LocalDLengthy,0)$) {$A_6'$};} 

	\draw pic [draw,->,angle radius=1.3cm, shift={(d)}] {angle = dB--d--e}; 
	
	}
	\pgfmathsetmacro\radius{30.7+5*\i}
	\draw pic [draw,->,angle radius=\radius, shift={(c)}] {angle = d--c--cB};
}

\end{tikzpicture}
    \caption{\textbf{a}: Sketch of a standard triple-axis instrument depicting the six angles, $A_1$-$A_6$. \textbf{b}: Zoom on multiplexing secondary spectrometer with two scattering angles and two final energies measured simultaneously.}
    \label{fig:TripleAxisOverview}
\end{figure}

Figure~\ref{fig:TripleAxisOverview}a shows a sketch of a typical triple-axis spectrometer with the six relevant rotation axes, $A_1$-$A_6$. The monochromator is rotated with respect to the direct beam by $A_1$. Through the Bragg scattering condition the sample is positioned at $A_2=2A_1$, such that a single incoming neutron wavevector, $\kkk_i$ (and thereby the energy $E_i$) is selected. The sample rotation is denoted $A_3$, while $A_4$ is the angle between the incident beam on the sample and the analyser. Similarly to the monochromator, the analyser is rotated by $A_5$ and the detector is placed at $A_6=2A_5$ to select a single final neutron wavevector, $\kkk_f$  (and thereby the energy $E_f$). The momentum and energy transfer between the neutron and the sample are given by $\QQQ=\kkk_i-\kkk_f$ and $\Delta E=E_i-E_f$, which can be calculated from $A_1-A_6$ and knowing the properties of the monochromator and analyser. We refer to Ref. \cite{TripleAxis} for a more extensive description of triple-axis spectrometers.

A multiplexing triple-axis instrument uses multiple analysers and detectors which leads to simultaneous measurements of the neutron scattering intensity at multiple $A_4-A_6$ positions (see Fig.~\ref{fig:TripleAxisOverview}b). In some instruments position sensitive detectors collect neutrons from multiple analysers. In addition, on most multiplexing instruments (including CAMEA) the scattering direction off the analyzers is rotated by 90$^\circ$ when compared to standard triple-axis instruments, thus scattering neutrons onto detectors that are out-of-plane. This impacts the resolution with which particular ($\QQQ$, $\Delta E$)-points are measured, but is mostly employed for practical reasons to allow for a maximal density of analyzer-detector pairs. The rotation out of plane changes the resolution calculation and renders current schemes inapplicable. Work is ongoing to account for this change.


\subsection{Coordinate systems}
\begin{figure}[ht!]
    \centering
    \includegraphics[height=0.53\linewidth]{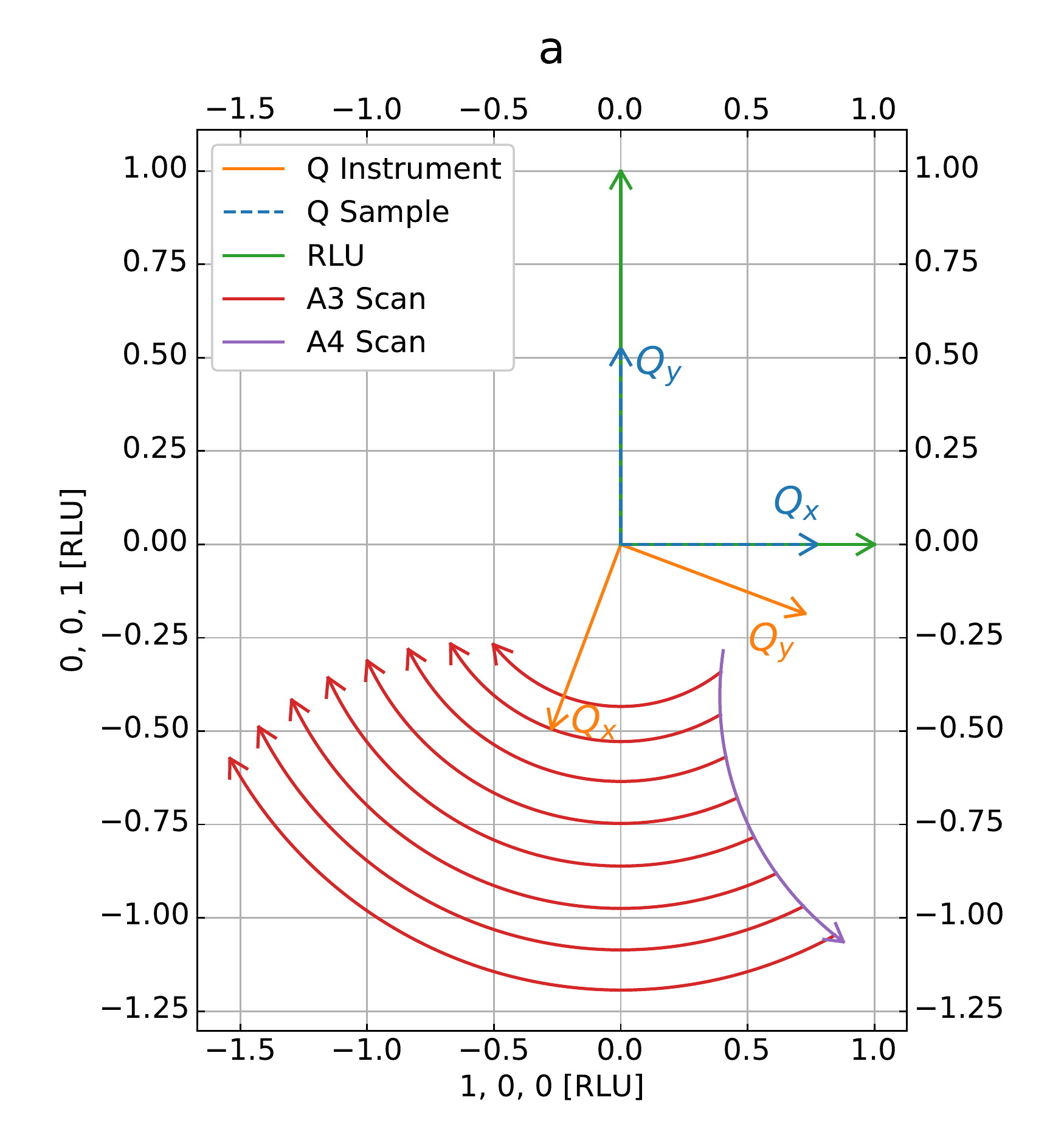}
    \includegraphics[height=0.53\linewidth]{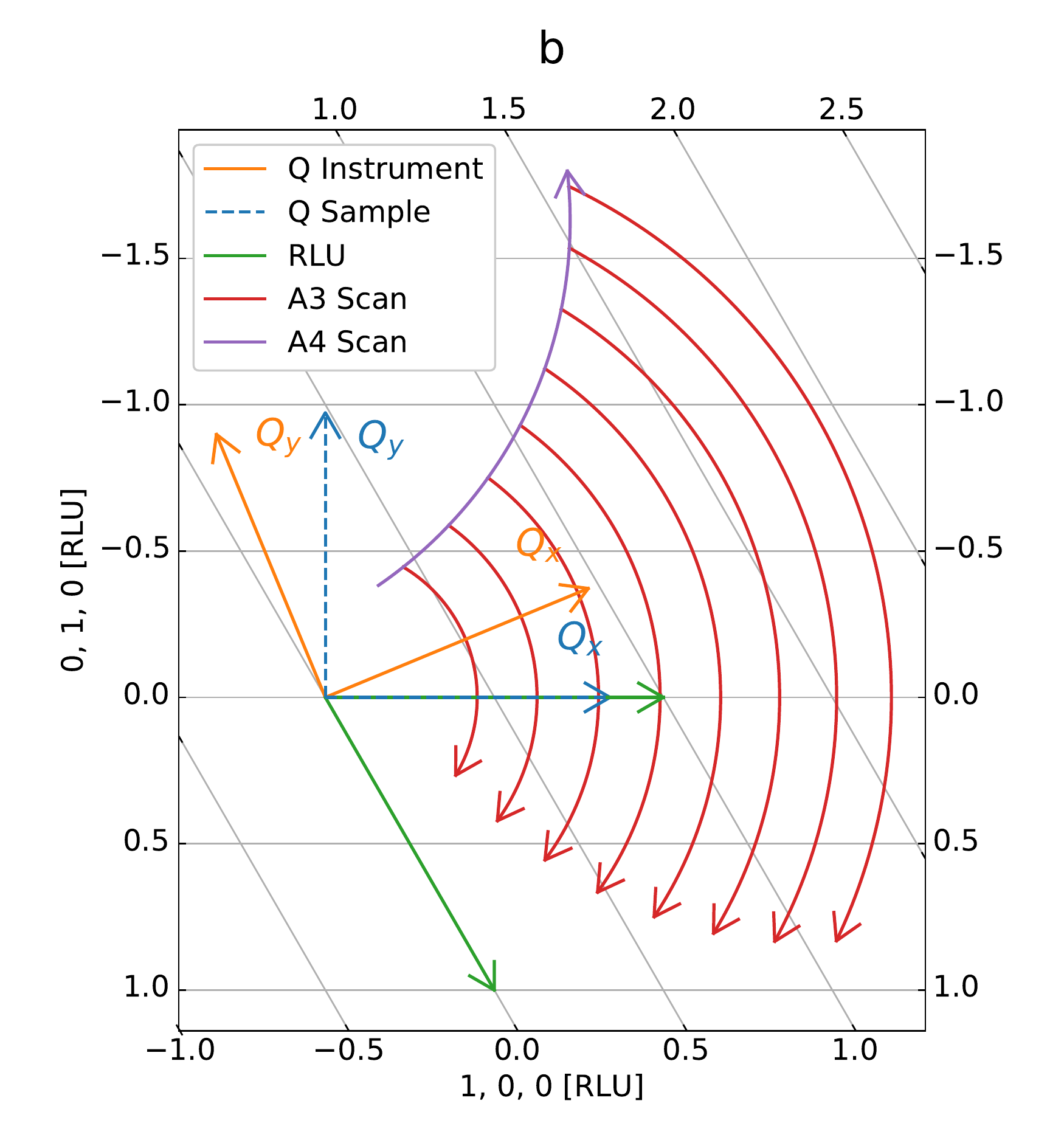}
    \caption{Overview of coordinate axes used for projection and RLU axes for \textbf{a}: tetragonal MnF$_2$ and \textbf{b}: hexagonal YMnO$_3$. The $A_3$ scan lines cover 90 $^\circ$ and have constant $A_4$ values between -16 $^\circ$ and -72 $^\circ$ covered by a single instrument setting by CAMEA.}
    \label{fig:Axes}
\end{figure}

In \MJOLNIR{}, the data treatment of multiplexing instruments is facilitated by defining several related coordinate systems that describe the scattering vector, $\QQQ$, illustrated in Fig.~\ref{fig:Axes}.

The instrument coordinate system is an orthonormal coordinate system that is sample independent. It is defined by $C_\text{instr}=(\QQQ_x,\QQQ_y,\QQQ_z,\Delta E)$, where $\QQQ_y$ is parallel to $\kkk_i$ and orthogonal to $\QQQ_x$ within the scattering plane. The out-of-plane component $\QQQ_z$ is zero in all cases.

The sample coordinate system shares the axes of instrument coordinate system, but is rotated so that the first axis points along a reciprocal lattice vector of the sample: $C_\text{sample} = (\tilde{\QQQ}_x,\tilde{\QQQ}_y,\tilde{\QQQ}_z,\Delta E)$. 

The reciprocal lattice coordinate system is defined by two reciprocal lattice vectors in the scattering plane, $C_\text{RLU}=(\PPP_1,\PPP_2,\Delta E)$, which is often chosen to be along high symmetry directions. For samples with cubic, tetragonal, or orthorhombic symmetry, the reciprocal lattice coordinate system overlaps with the sample coordinate system, such as illustrated in Fig.~\ref{fig:Axes}a. In contrast, the reciprocal lattice vectors are not orthorgonal in hexagonal, trigonal, monoclinic or triclinic systems, see Fig.~\ref{fig:Axes}b.

A typical measurement on CAMEA consists of rotating $A_3$ while keeping all other angles constant. This generates data points as shown in Fig.~\ref{fig:Axes} by the red lines.

\section{The Scope of \MJOLNIR}

\MJOLNIR{} is designed to provide tools for three purposes: 1) conversion of data into the reciprocal lattice coordinate system or the sample coordinate system. 2) Normalisation and visualisation of the data in 1, 2 and 3 dimensions, and 3) analysis of the data. We now describe the scope of \MJOLNIR{} in these three aspects.
 
\subsection{Data conversion}

Conversion of data to the reciprocal lattice coordinate system requires detailed knowledge of the specific instrument setup. \MJOLNIR{} has initially been developed for CAMEA, and thus includes an accurate description of this instrument. We, however, aim for \MJOLNIR{} to be used across many multiplexing spectrometers such as MultiFLEXX\cite{Groitl2017}, FlatCone\cite{FlatCone} and Bambus\cite{Lim2014}. We have thus developed tools to model any multiplexing triple-axis type instrument in \MJOLNIR{}. We note that MultiFLEXX is being moved to the Heinz Maier-Leibnitz Zentrum due to limited activities at the Helmholtz-Zentrum in Berlin.
 
There are two methods by which the required instrument parameters can be extracted for the data conversion, which are explained in detail in Ref.~\cite{Lass2020CAMEA}. The parameters can either be calculated from the instrument design directly, assuming a perfect setup, or measured through an experiment allowing for small deviations. A direct measurement yields more accurate results, but requires some extra modelling. At CAMEA, the scattering angles are calculated, while final energies are measured directly. At MultiFLEXX both parameters are calculated\cite{Groitl2017}. \MJOLNIR{} supports both options.

Further differences among various multiplexing instruments arise from the fact that some of them employ a prismatic analyser concept of several energies scattered from one analyzer \cite{Birk2014}, while others do not. However, by implementing the prismatic concept, the standard setup will be a limiting case of just a single energy, and is thus also supported.

Along these lines, the requirements for the upcoming indirect time-of-flight spectrometer BIFROST at the European Spallation Source are worth noting \cite{CAMEAProposal,Freeman2015}. The instrument shares many similarities with CAMEA; both possess multiplexing secondary spectrometers, utilising the prismatic concept in combination with position sensitive detectors. However, their primary spectrometers differ. CAMEA is situated at a quasi-continuous source using a monochromator to determine the incoming energy. BIFROST, on the other hand, will be situated on a pulsed source, determining the incoming energy by time-of-flight. In effect, BIFROST performs a scan over incoming energies within a defined window, whereas CAMEA employs one fixed incoming energy only. In practice this extends BIFROST data by an additional dimension when compared to CAMEA. We expect that extensions of \MJOLNIR{} will be suitable for handling data from BIFROST.

\subsection{Visualisation}
\MJOLNIR{} is designed to work directly with raw data, where each data point consists of a neutron count, a monitor count, and normalization value, as well as a position in reciprocal space. Acquiring multiple scans at almost identical positions will increase the likelihood that multiple points are in close vicinity of each other. One may argue that if their difference is smaller than the expected instrumental uncertainty they should be binned. The approach, however, imposes a resolution estimate reducing the transparency of the data process, as two points within a given distance in a certain parameter space might be binned in some part of the measurement volume but not in other parts. This is due to the fact that the instrument resolution changes across the probed volume of reciprocal space. Instead, we prefer to treat all data points separate as long as no user-defined tolerance is provided. 

\MJOLNIR{} provides tools for visualising data from any multiplexing instrument in 1, 2 and 3 dimensions in the reciprocal lattice coordinate system and in the sample coordinate system. For these purposes, the user defines the size of the bins as well as the directions of the desired cuts. \MJOLNIR{} then bins and normalises the raw data to the monitor count, and plots the binned data. The tools are described in detail in the documentation\cite{PSICAMEADataTreatment}.

We note that \MJOLNIR{} has been written for multiplexing triple-axis instruments specifically. The software is not suitable for data acquired on standard triple-axis instruments where only a one-dimensional sub-space of reciprocal space is measured. Conversely, direct geometry time-of-flight spectrometers produce 4-dimensional data sets, whose handeling is outside the scope of \MJOLNIR{}.

\subsection{Data analysis}
Because all data is kept in the original un-tampered form throughout the data treatment process, different fitting algorithms and methods can be employed. This is important for non-standard Poisson and Multinomial fitting routines\cite{lass2020Statistics}, where it is important to keep integer neutron counts as opposed to normalized counts. Currently, fitting is supported through a series of 1D cuts in reciprocal space for constant energy, or constant $\QQQ$. At present, interface tools with the uFit\cite{ufit} program is under development and will in the long run also interface with analysis software, such as SpinW\cite{Toth2015}.

\section{Program Structure} 
The \MJOLNIR{} package is composed of several modules, each dedicated to a specific task. 
A virtual model of the instrument is generated using the \code{Geometry} module. The actual data processing and visualization objects are located in the \code{DataSet} module. The  data fitting process is kept separate from the data objects in the \code{Statistics} module in an effort to clearly separate data conversion and data analysis, and to be also able to switch fitting routines and software. At the time of writing this module is being merged with the capabilities of uFit\cite{ufit}, although these efforts are not reported in the following. General functions and repeatedly used subroutines are located in the \code{_tools} module. One of them is a collection of triple-axis conversion commands called \code{TasUBLibDEG}, which was translated from the \CC library TasUBlib based on Lumsden $et.$\ $al.$\ \cite{Lumsden2005}. In the following sections, we highlight some key features of the different modules.

\subsection{Geometry}
The objects and methods in the \code{Geometry} module have two principal purposes; the first is the virtual representation of the instruments, the second is to facilitate the generation of normalisation tables in which the calibrated pixel efficiency, final energy, and scattering angles of the detectors are stored. The normalisation has been separated from the implementation of the data structure in an effort of generalisation, and to signify its connection with the instrument as opposed to the data treatment. A virtual representation of the instrument will be particularly useful in foreseen future features of the data treatment such as calculations of the resolution function.

\paragraph{Virtual Instrument}
The creation of a virtual instrument is done using the \code{Instrument} object and the subsequent \code{Wedge}, \code{Analyser}, and \code{Detector} objects. These encode their real world counterparts, containing information about instrument relevant positions, directions, and $d$-spacings. The simulation of an instrument is implemented either via a script that adds the objects to the instrument structure, or by means of an XML file.

Different instruments feature varying setups. In the simplest case, there is a one-to-one correspondence between detectors and analyser. This is the case for MultiFLEXX as each detector corresponds to a single analyser. The situation is more involved at CAMEA as the instrument consists of 13 position sensitive detectors inside a 8$^\circ$ wide wedge. Each detector detects neutrons scattered from 8 different analysers, typically corresponding to 24 to 64 different energies, when the prismatic analyzer concept is employed. In such cases where a detector receives neutrons from different analysers, a position sensitive detector with a defined pixel numbering is simulated. This allows splitting of detectors into sub-parts, identifying neutrons from the different analysers.

\subsection{Data}
All code that is connected to the data conversion and data treatment is grouped in the \code{Data} module. This includes the \code{DataFile} and \code{Sample} objects, which refer to instrument and sample parameters of individual scan files. This, for instance, enables masking options for specified data regions in individual data files or across an entire data set.

Experiments on multiplexing instruments often produce a number of data files with similar parameters that can be combined into a single data set. For this reason we created the \code{DataSet} object that deals with multiple data files simultaneously. It represents an abstraction of a list containing \code{DataFile} objects, and hosts a number of methods and helper functions to perform cuts and plots. 

\paragraph{Conversion}
Once the experimental geometry and all instrument settings are known, the conversion from the detector positions to reciprocal lattice units happens in two steps. Fist, the $A_4$ angles and final energies of the instrument reference frame are transformed into the instrument coordinate system, and then to the reciprocal lattice unit system of the sample. Both conversions are based on the formalism of TasUBlib \cite{Lumsden2005}. 

Most often, a data file arises by scanning one or more instruments settings, typically the incoming energy, the sample rotation, or the rotation angle of the analyzer-detector tank that impacts the scattering angle of all detectors. However, \MJOLNIR{} allows for the most general case of multiple changing parameters, which happens when scanning both the sample rotation and scattering angle simultaneously. 

Data from each file is converted to a structure of shape $n_p\times n_\text{detectors}\times n_\text{pixels}$, where $n_p$, $n_\text{detectors}$ and $n_\text{pixels}$ is the number of scan points, detectors and pixels per detector, respectively. One such object is created for $A_3$, $A_4$, $\Delta E$, $qx$, $qy$, $H$, $K$, $L$, the normalisation, monitor and neutron count $I$. Here, normalization refers to the combined analyzer and detector efficiency, while the monitor count keeps track of the neutron count in the guide monitor. The converted data are the basis for any further process. We note that the normalisation, monitor and $I$ are not combined, but kept separately.

\paragraph{Performing cuts}
Directly visualizing a complete scan file as a point cloud in 3D reciprocal space is often disadvantageous. Instead, cuts along specific lines or planes in reciprocal space are performed to highlight specific features. Due to the usual scarcity of regions of interest it may also be preferable to avoid cuts where only low scattering areas are present. Multiple cutting and corresponding plotting methods have been implemented in \MJOLNIR{}. These include 1D cuts along a constant energy or momentum transfer line and 2D intensity maps connecting multiple scattering points. These cuts are useful to visualise the data, and are also the basis for further data processing such as fitting. Further information on the visualising routine is given below.

\paragraph{Masking}
As CAMEA measures a large region in reciprocal space during a single scan, additional unintended signals are sometimes detected. These may be extrinsic features such as spurious peaks or additional phonons or magnetic contributions that do not concern the addressed scientific question. It may be preferential to mask these contributions, before cuts or fits are preformed. In \MJOLNIR{}, this is supported by the \code{Mask} module where different options are available. They can be combined using standard Boolean algebraic operations, which are described in detail in the \MJOLNIR{} documentation\cite{PSICAMEADataTreatment}.

\subsection{Statistics and Fitting}\label{sec:Fitting}
The main goal of any neutron scattering experiment is to gain deeper insight into microscopic parameters of the studied material. This happens either through direct determination of microscopic properties or indirectly through a comparison with theoretical models. In both cases some kind of fitting is needed. \MJOLNIR{} provides a statistical tool, which allows for different statistical approaches. In particular when dealing with Poisson statistics, the normalisation and neutron intensity at each measured reciprocal lattice  needs to be taken into account \cite{lass2020Statistics}. An accurate fitting routine of such data is currently developed by merging \MJOLNIR{} with the existing software package of uFit. We refer to the documentation of that uFit for details\cite{ufit}.

\subsection{Interfaces}
Three different interfaces have been created which serve different purposes, also providing different levels of control. The main interface is the scripting interface of \MJOLNIR{} which is imported in a python script. It supports all features of the software and allows creating new virtual instruments, normalization tables, and to convert and visualise the data. The scripting interface is the backbone of the software. A limited command line interface can be used to shortcut some of the key functions. These include reporting, conversion and visualization functions, serving as a quick way of checking data during experiments. Third, we have also developed a graphical user interface meant to both allow quick figure generation through a point-and-click interface and to also guide users towards the scripting interface.

\begin{figure}[ht]
    \centering
    \includegraphics[width=0.8\linewidth]{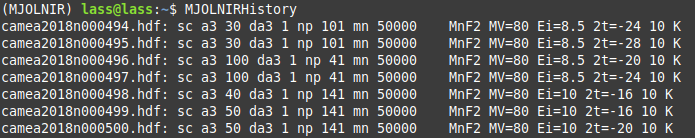}
    \caption{Output of \code{MJOLNIRHistory} for data files 494 to 500 taken in 2018, showing scan command and title ("MnF2 MV=80 Ei=10 2t=-20 10 K").}
    \label{fig:CommandLine}
\end{figure}

\paragraph{Command Line Interface}
Four commands have been created that act as a short list for \code{MJOLNIR} scripts. These are the \code{MJOLNIR3DView}-method plotting the interactive View3D window, the \code{MJOLNIRConvert}-method converting data sets from HDF to NXS using a specified binning, the \code{MJOLNIRCalibrationInspector}-method that displays the current normalization options, $A_4$ values and final energies, and the \code{MJOLNIRHistory}-method that reports the data set content. In each case the respective data files are entered directly or through a file dialog. Further variables or input parameters can be attached to control properties such as the bin sizes used in conversion or the type of plot that is generated. An example of the \code{MJOLNIRHistory}-method is shown in Fig.~\ref{fig:CommandLine}.

\paragraph{Graphical User Interface}
A user-friendly interface has been created for users that prefer to avoid using Python in a scripting-based environment. The interface is built directly on top of the \MJOLNIR{} package and provides access to a variety of its key features. The tool can generate different plots and inspect experimental data on the fly. We also implemented the possibility to directly generate Python scripts from the GUI interface. This is useful if the user prefers to add customary modifications to figures, for instance. Adhering to the desire of being usable across many operating systems, we wrote \code{MJOLNIRGui} using the python bindings of Qt\cite{Qt}. In combination with FBS package\cite{fbs} it allows for the creation of installers across different operating systems.

\begin{figure}[ht!]
    \centering
    \includegraphics[width=0.9\linewidth]{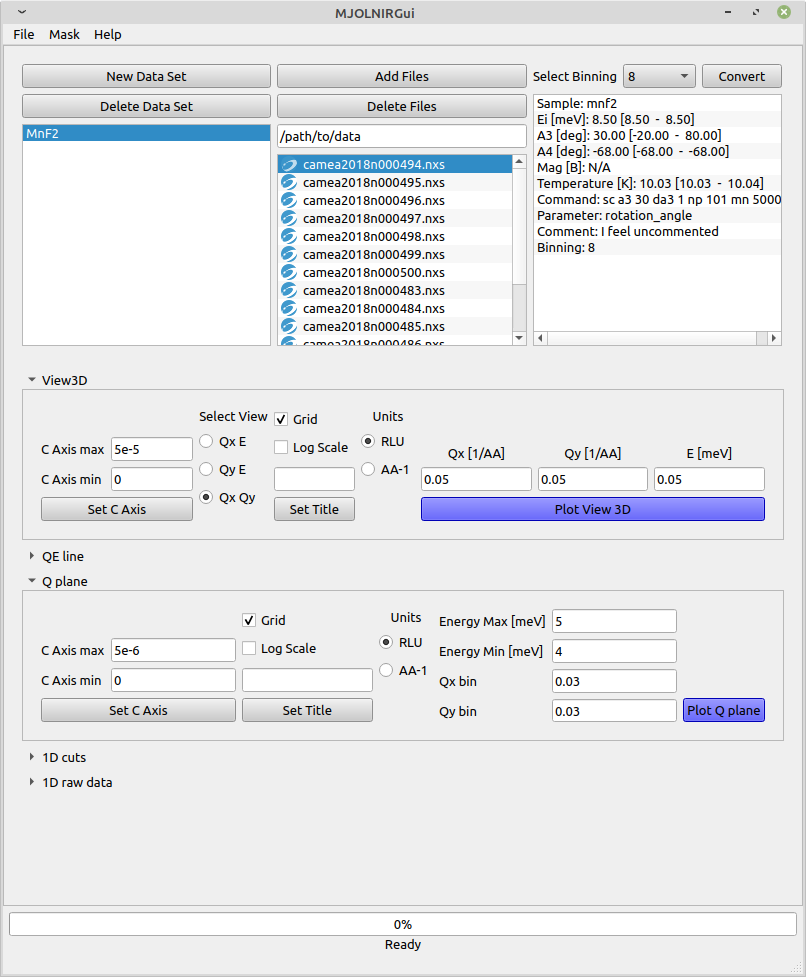}
    \caption{Overview of the graphical user interface of \MJOLNIR{} denoted \code{MJOLNIRGui}, where blue coloured buttons suggest next steps of the data treatment. Selecting a data file in a data set shows its properties in the right upper-most window.}
    \label{fig:MJOLNIRGui}
\end{figure}

\section{Using \MJOLNIR{}}
In this section we show examples of how \MJOLNIR{} can be used during an experiment. The data and a tutorial replicating the figures produced here can be found in the documentation of \MJOLNIR{}\cite{PSICAMEADataTreatment}. The presented spin waves were acquired on a large MnF$_2$ single crystal at $T$= 10 K acquired at the CAMEA instrument at PSI. Further details are found in the CAMEA commissioning article\cite{Lass2020CAMEA}.

\subsection{Loading and converting data}

Using the \MJOLNIRGui{}, we first create a new Data set, then add the data files, load them into \MJOLNIR{}, and finally convert the data to the reciprocal lattice space. These steps are completed within the GUI, or using the scripting interface.

\subsection{Overview of data: Viewer3D}
For a quick overview of the data, the Viewer3D method is particularly useful. The method bins and plots the data in the instrument coordinate system $(\QQQ_x,\QQQ_y,\Delta E)$ or the reciprocal lattice coordinate system, $(\PPP_1,\PPP_2,\Delta E$). 
The data is binned via the SciPy \code{HistogramDD} into equi-sized voxels, whose size can be specified by the user. 
The main reason for utilising a constant binning is rapid processing time. We chose to perform the binning in the orthonormal sample coordinate system to truthfully visualise the data, which would not be the case if it was performed in a geometry where the reciprocal lattice basis vectors had different lengths. Alternatively, one could plot constant energy planes where in a polar coordinate system. This is because an $A_3$ rotation represents a rotation around the origin, which yields almost equally distributed data points as a function of $|Q|$. We implemented this option into the \code{plotQPlane} method.

The \code{View3D} module comes with an interactive GUI that allows the user to step through the data, plane by plane, along three different directions, 
\textit{i.e.}, along the two projection vectors and the energy. It is possible to perform cuts along arbitrary directions, for which more computational power is required. Examples of the output of View3D are shown in Fig.~\ref{fig:Viewer3DSequence}, and the script needed to produce this figure is given in appendix \ref{app:viewer3D}, as generated by the Gui interface.

\begin{figure*}[ht]
    \centering
    \includegraphics[width=0.49\linewidth]{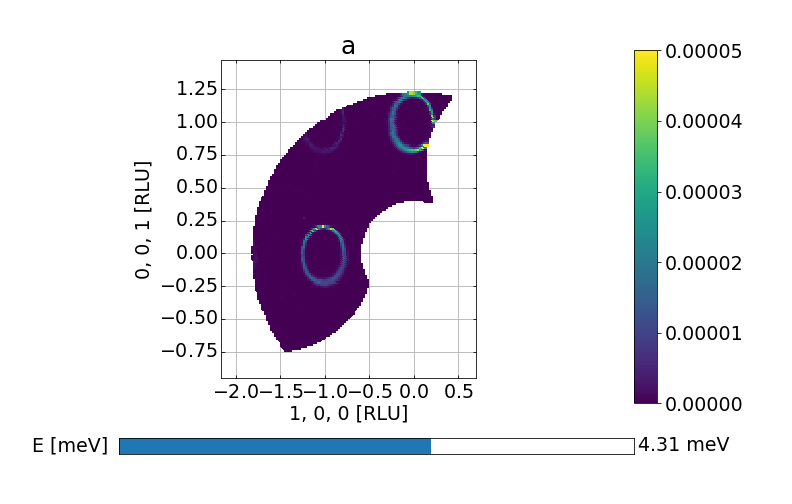}
    \includegraphics[width=0.49\linewidth]{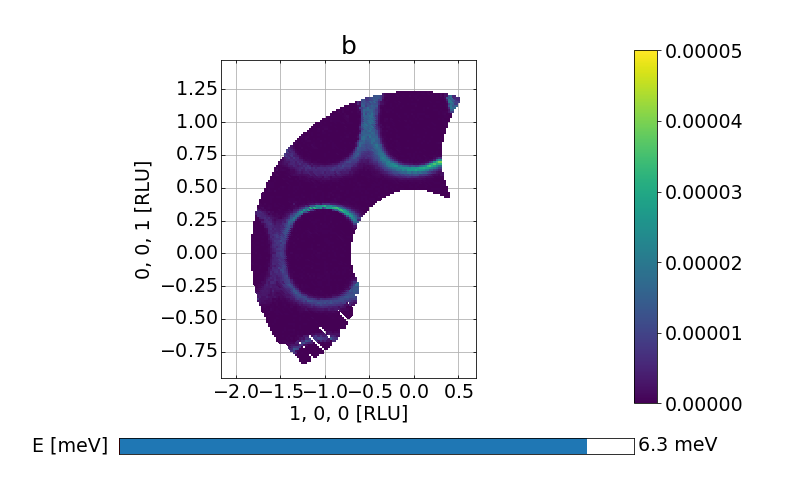}
    \includegraphics[width=0.49\linewidth]{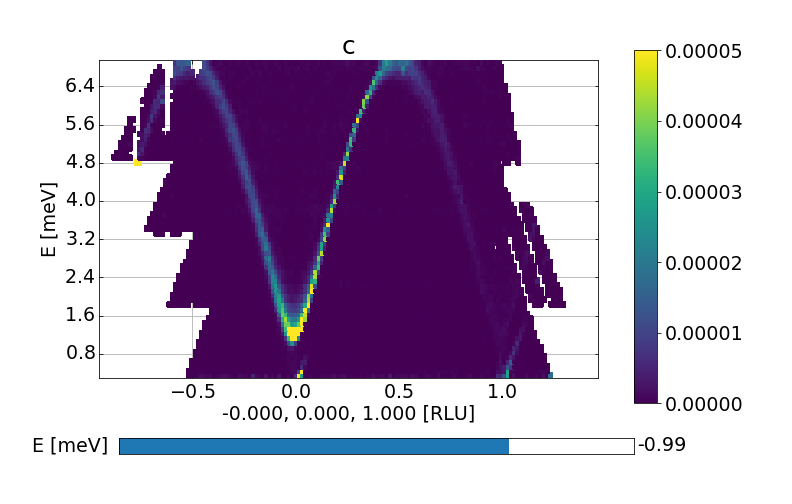}
    \includegraphics[width=0.49\linewidth]{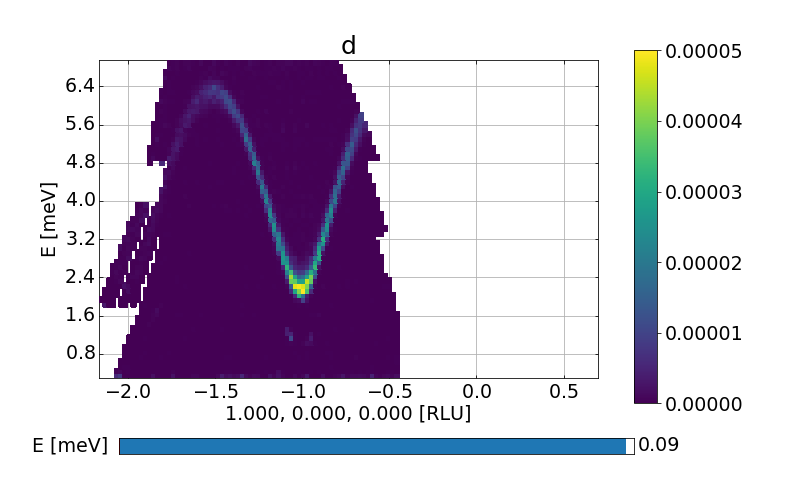}
    \caption{Series of pictures from \code{Viewer3D} showing data from MnF$_2$ for 2 different constant-$E$ planes in Figs.~a and b and two $(q,E)$ cuts along the principal directions in Figs.~c and d.}
    \label{fig:Viewer3DSequence}
\end{figure*}

\subsection{Cuts through the data: plotCutQELine}
The data may be understood deeper when plotting the intensity as function of energy and $\QQQ$ along a path in reciprocal space. The \code{plotCutQELine}-module has been developed for this purpose. 
The code is composed of smaller building blocks, ensuring a successful visualisation of arbitrary two-dimensional cuts with different binning sizes and relative distances in reciprocal space.

\begin{figure*}[ht]
    \centering
    \includegraphics[width=0.49\linewidth]{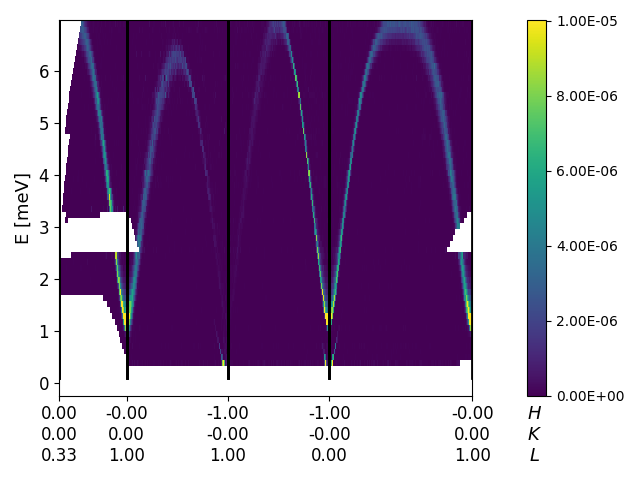}
    \includegraphics[width=0.49\linewidth]{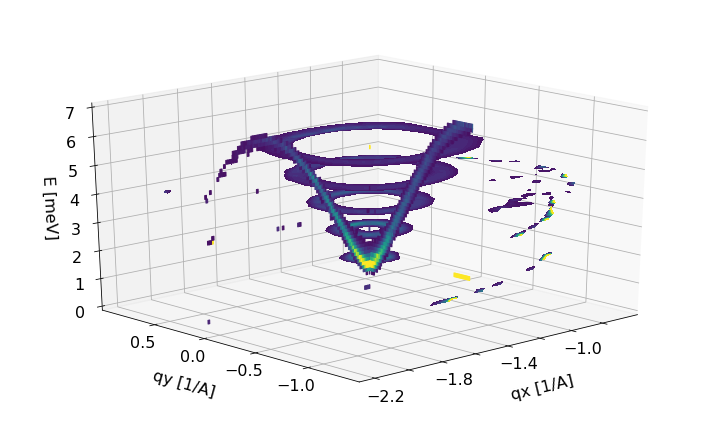}
    \caption{\textbf{a}: Intensity plotted as function of energy and $\QQQ$ for MnF$_2$ using \code{plotCutQELine} and 5 $\QQQ$ points. \textbf{b}: 3D visualisation of scattering intensity for MnF$_2$. Transparent parts signify low intensity.}
    \label{fig:plotCutQELineAnd3D}
\end{figure*}

Each cut unifies a collection of one-dimensional cuts of a certain Q-width and constant energy window. An example is shown in Fig.~\ref{fig:plotCutQELineAnd3D}a using 5 different $Q$ points of MnF$_2$. 
A mouse-over function displays the normalized intensity, the central relative lattice position, the normalization, monitor and the number of binning points to create the shown pixel intensity. 

The plotting method also allows to display 2D planes embedded in a three-dimensional figure as shown in Fig.~\ref{fig:plotCutQELineAnd3D}b. The plots have been combined with constant energy plots generated by the \code{plotQPlane} method, and the colour map is chosen such that low intensity points are transparent.

\FloatBarrier{}

\section{Conclusion}
We have developed the \MJOLNIR{}~software package to convert, visualize and analyse data from multiplexing triple-axis instruments. It has already been successfully used during the commissioning of CAMEA at PSI, and the analysis of the collected data thereof. The package will be further extended in the future to include direct support of other multiplexing instruments, prediction of  spurions, experimental planning tools, advanced fitting tools, and resolution and absorption calculations. We anticipate \MJOLNIR{} to greatly benefit users of multiplexing triple-axis instruments.

\section{Conflict of Interest}
We wish to confirm that there are no known conflicts of interest associated with this publication.

\section{Funding}

This research was founded by the Danish Agency for Research and Innovation through DanScatt grant 7055-00007B. J.L. was supported by Nordforsk project 81695: NNSP School, and by the Paul Scherrer Institut.

HJ was supported by the European Union's Horizon 2020 research and innovation program under the Marie Skłodowska-Curie grant agreement No 701647.

\section*{Acknowledgements}
Many people have partaken in discussion concerning the software requirements and development. We would especially like to thank Christof Niedermayer and Jonas Okkels Birk for many discussion early in the development process. This laid much of the foundation for its formulation. Also Paul Steffens and Siqin Ming have participated in many fruitful discussions. We especially acknowledge them for their insights into Voronoi methods and adaptive binning. We also thank the people standing in as guinea-pigs for the first versions of the software. This includes Sofie Janas, Morten Lunn Haubro, Virgile Favre, Luc Testa, and Stephan Allenspach.


\bibliography{bibliography}
\appendix
\newpage

\section{Viewer3D script}\label{app:viewer3D}
The following script was generated by \MJOLNIRGui\ using the MnF$_2$ data.

\begin{lstlisting}[language=Python,basicstyle=\scriptsize]
import matplotlib.pyplot as plt
try:
    import IPython
    shell = IPython.get_ipython()
    shell.enable_matplotlib(gui='qt')
except:
    pass


from MJOLNIR import _tools
from MJOLNIR.Data import DataSet

import numpy as np
from os import path
dataFiles = _tools.fileListGenerator("483-489,494-500",r"path/to/data",2018)

MnF2 = DataSet.DataSet(dataFiles)
# Run the converter. This automatically generates nxs-file(s). 
# Binning can be changed with binning argument.
MnF2.convertDataFile(binning = 8,saveFile=False)


# Plotting data quickly in equi-sized voxels can be done by
Viewer = MnF2.View3D(0.05,0.05,0.05, grid=True)
# Above, all data is binned in voxels of size 0.05/AA, 0.05/AA, and 0.05 meV.
# Automatically, data is plotted in reciprocal lattice as provided by the
# UB matrix saved in the data file. Alternatively, one can plot in 'raw'
# coordinates (orthogonal and in units of 1/AA) by issuing rlu=False above.

Viewer.caxis=(0,5e-5)
# Without any intervention data is usually plotted on a useless colour scale.
# This is countered by specifying min and max for colour by the call above.
# Alternatively, one can provide this as input to View3D

plt.show()
\end{lstlisting}

\end{document}